\documentclass[%
reprint,
superscriptaddress,
nofootinbib,
nobibnotes,
 amsmath,amssymb,
 aps,
 prl,
 longbibliography,
 twocolumn,
]{revtex4-2}% 4-2 has problem with bibtex

\usepackage{epic}
\usepackage{float}
\usepackage[percent]{overpic}
\usepackage{graphicx, comment}% Include figure files
\usepackage{dcolumn}% Align table columns on decimal point
\usepackage{bm}% bold math
\usepackage{xcolor}

\usepackage{hyperref}% add hypertext capabilities
\hypersetup{
    colorlinks=true,
    linkcolor=blue,
    filecolor=magenta,      
    urlcolor=cyan,
    pdftitle={uboone_NCpi0PRL},
    pdfpagemode=FullScreen,
    }

\usepackage[normalem]{ulem}
\usepackage{amsfonts}
\usepackage{url}
\usepackage{graphicx,subcaption}% Include figure files
\usepackage[newcommands]{ragged2e}

\DeclareCaptionJustification{justified}{\justifying}
\usepackage{verbatim}%for multiline comments

\begin{document}
%\linenumbers
\setlength{\parskip}{0pt}

\preprint{APS/123-QED}

% List of institutions in command form:
\newcommand{\ANL}{Argonne National Laboratory (ANL), Lemont, IL, 60439, USA}
\newcommand{\Bern}{Universit{\"a}t Bern, Bern CH-3012, Switzerland}
\newcommand{\BNL}{Brookhaven National Laboratory (BNL), Upton, NY, 11973, USA}
\newcommand{\UCSB}{University of California, Santa Barbara, CA, 93106, USA}
\newcommand{\Cambridge}{University of Cambridge, Cambridge CB3 0HE, United Kingdom}
\newcommand{\CIEMAT}{Centro de Investigaciones Energ\'{e}ticas, Medioambientales y Tecnol\'{o}gicas (CIEMAT), Madrid E-28040, Spain}
\newcommand{\Chicago}{University of Chicago, Chicago, IL, 60637, USA}
\newcommand{\Cincinnati}{University of Cincinnati, Cincinnati, OH, 45221, USA}
\newcommand{\CSU}{Colorado State University, Fort Collins, CO, 80523, USA}
\newcommand{\Columbia}{Columbia University, New York, NY, 10027, USA}
\newcommand{\Edinburgh}{University of Edinburgh, Edinburgh EH9 3FD, United Kingdom}
\newcommand{\FNAL}{Fermi National Accelerator Laboratory (FNAL), Batavia, IL 60510, USA}
\newcommand{\Granada}{Universidad de Granada, Granada E-18071, Spain}
\newcommand{\Harvard}{Harvard University, Cambridge, MA 02138, USA}
\newcommand{\IIT}{Illinois Institute of Technology (IIT), Chicago, IL 60616, USA}
\newcommand{\Indiana}{Indiana University, Bloomington, IN 47405, USA}
\newcommand{\KSU}{Kansas State University (KSU), Manhattan, KS, 66506, USA}
\newcommand{\Lancaster}{Lancaster University, Lancaster LA1 4YW, United Kingdom}
\newcommand{\LANL}{Los Alamos National Laboratory (LANL), Los Alamos, NM, 87545, USA}
\newcommand{\Louisiana}{Louisiana State University, Baton Rouge, LA, 70803, USA}
\newcommand{\Manchester}{The University of Manchester, Manchester M13 9PL, United Kingdom}
\newcommand{\MIT}{Massachusetts Institute of Technology (MIT), Cambridge, MA, 02139, USA}
\newcommand{\Michigan}{University of Michigan, Ann Arbor, MI, 48109, USA}
\newcommand{\MSU}{Michigan State University, East Lansing, MI 48824, USA}
\newcommand{\Minnesota}{University of Minnesota, Minneapolis, MN, 55455, USA}
\newcommand{\Nankai}{Nankai University, Nankai District, Tianjin 300071, China}
\newcommand{\NMSU}{New Mexico State University (NMSU), Las Cruces, NM, 88003, USA}
\newcommand{\Oxford}{University of Oxford, Oxford OX1 3RH, United Kingdom}
\newcommand{\Pitt}{University of Pittsburgh, Pittsburgh, PA, 15260, USA}
\newcommand{\Rutgers}{Rutgers University, Piscataway, NJ, 08854, USA}
\newcommand{\SLAC}{SLAC National Accelerator Laboratory, Menlo Park, CA, 94025, USA}
\newcommand{\SDSMT}{South Dakota School of Mines and Technology (SDSMT), Rapid City, SD, 57701, USA}
\newcommand{\Maine}{University of Southern Maine, Portland, ME, 04104, USA}
\newcommand{\Syracuse}{Syracuse University, Syracuse, NY, 13244, USA}
\newcommand{\TelAviv}{Tel Aviv University, Tel Aviv, Israel, 69978}
\newcommand{\Tennessee}{University of Tennessee, Knoxville, TN, 37996, USA}
\newcommand{\UTA}{University of Texas, Arlington, TX, 76019, USA}
\newcommand{\Tufts}{Tufts University, Medford, MA, 02155, USA}
\newcommand{\UCL}{University College London, London WC1E 6BT, United Kingdom}
\newcommand{\VTech}{Center for Neutrino Physics, Virginia Tech, Blacksburg, VA, 24061, USA}
\newcommand{\Warwick}{University of Warwick, Coventry CV4 7AL, United Kingdom}
\newcommand{\Yale}{Wright Laboratory, Department of Physics, Yale University, New Haven, CT, 06520, USA}
%%\newcommand{\listerThanks}{Now at University of Wisconsin, Madison}

% So that institutions appear in alphabetical order:
\affiliation{\ANL}
\affiliation{\Bern}
\affiliation{\BNL}
\affiliation{\UCSB}
\affiliation{\Cambridge}
\affiliation{\CIEMAT}
\affiliation{\Chicago}
\affiliation{\Cincinnati}
\affiliation{\CSU}
\affiliation{\Columbia}
\affiliation{\Edinburgh}
\affiliation{\FNAL}
\affiliation{\Granada}
\affiliation{\Harvard}
\affiliation{\IIT}
\affiliation{\Indiana}
\affiliation{\KSU}
\affiliation{\Lancaster}
\affiliation{\LANL}
\affiliation{\Louisiana}
\affiliation{\Manchester}
\affiliation{\MIT}
\affiliation{\Michigan}
\affiliation{\MSU}
\affiliation{\Minnesota}
\affiliation{\Nankai}
\affiliation{\NMSU}
\affiliation{\Oxford}
\affiliation{\Pitt}
\affiliation{\Rutgers}
\affiliation{\SLAC}
\affiliation{\SDSMT}
\affiliation{\Maine}
\affiliation{\Syracuse}
\affiliation{\TelAviv}
\affiliation{\Tennessee}
\affiliation{\UTA}
\affiliation{\Tufts}
\affiliation{\UCL}
\affiliation{\VTech}
\affiliation{\Warwick}
\affiliation{\Yale}

% Authors in alphabetical order
\author{P.~Abratenko} \affiliation{\Tufts}
\author{O.~Alterkait} \affiliation{\Tufts}
\author{D.~Andrade~Aldana} \affiliation{\IIT}
\author{L.~Arellano} \affiliation{\Manchester}
\author{J.~Asaadi} \affiliation{\UTA}
\author{A.~Ashkenazi}\affiliation{\TelAviv}
\author{S.~Balasubramanian}\affiliation{\FNAL}
\author{B.~Baller} \affiliation{\FNAL}
\author{A.~Barnard} \affiliation{\Oxford}
\author{G.~Barr} \affiliation{\Oxford}
% ...

\collaboration{The MicroBooNE Collaboration}
\thanks{microboone\_info@fnal.gov}\noaffiliation

%\date{\today}
\hyphenation{Micro-BooNE}
\title{First double-differential cross section measurement of neutral-current $\pi^0$ production in neutrino-argon scattering in the MicroBooNE detector}

\begin{abstract}
%\begin{linenumbers}
We report the first double-differential cross section measurement of neutral-current neutral pion (NC$\pi^0$) production in neutrino-argon scattering, as well as single-differential measurements of the same channel in terms of final states with and without protons. The kinematic variables of interest for these measurements are the $\pi^0$ momentum and the $\pi^0$ scattering angle with respect to the neutrino beam. A total of 4971 candidate NC$\pi^0$ events fully-contained within the MicroBooNE detector are selected using data collected at a mean neutrino energy of $\sim 0.8$~GeV from $6.4\times10^{20}$ protons on target from the Booster Neutrino Beam at the Fermi National Accelerator Laboratory. After extensive data-driven model validation to ensure unbiased unfolding, the Wiener-SVD method is used to extract nominal flux-averaged cross sections. The results are compared to predictions from commonly used neutrino event generators, which tend to overpredict the measured NC$\pi^0$ cross section, especially in the 0.2-0.5~GeV/c $\pi^0$ momentum range and at forward scattering angles. Events with at least one proton present in the final state are also underestimated. This data will help improve the modeling of NC$\pi^0$ production, which represents a major background in measurements of charge-parity violation in the neutrino sector and in searches for new physics beyond the Standard Model.
%\end{linenumbers}
\end{abstract}

\maketitle

Modern accelerator-based neutrino experiments are capable of expansive physics programs that address a variety of important topics. These include charge-parity violation in the neutrino sector~\cite{cp_t2k,cp}, the neutrino mass ordering~\cite{Qian:2015waa}, measurements of rare Standard Model processes~\cite{glee,nomad_gamma,t2k_ncdelta,etta}, searches for sterile neutrinos~\cite{steriles1,steriles2} and other physics beyond the standard model (BSM)~\cite{BSM1,BSM5}. Many of these analyses require measuring the rate of interactions that produce single electrons~\cite{wc_osc,mb_sterile,t2k_steriles,nova_steriles,minos_steriles}, single photons~\cite{glee,t2k_ncdelta}, or boosted and overlapping $e^+e^-$ or $\gamma\gamma$ pairs~\cite{BSM2,BSM3,BSM4,BSM6} by selecting events that leave an electromagnetic shower signature in the detector. In the few-GeV neutrino energy regime relevant to these experiments, neutral-current neutral pion (NC$\pi^0$) production represents the primary background in single-shower selections. 

Below neutrino energies of about $1.5$~GeV, the NC$\pi^0$ channel is dominated by resonance interactions ~\cite{RS,RES2,pionFSI,res} where the initial neutrino-nucleon scattering produces a $\Delta$(1232) baryon that can decay to a nucleon and a $\pi^0$ that exit the nucleus. Coherent scattering~\cite{cohRS}, where the neutrino interacts with the nucleus as a whole rather than an individual nucleon, and final state interactions (FSI) experienced by hadrons produced through other interaction modes~\cite{pionFSI2,pionFSI3} also contribute to $\pi^0$ production. These processes are sub-dominant yet important in a robust description of the NC$\pi^0$ channel. 

Outside the nucleus, the $\pi^0$ decays to two photons with a 99\% branching ratio, resulting in a two shower topology. If one of these photons is not reconstructed, the NC$\pi^0$ event will be misidentified as a single-shower event leading to their prominence in single-shower selections. Precise theoretical modeling of NC$\pi^0$ production is thus needed to maximize the physics reach of neutrino experiments. This requires the support of detailed NC$\pi^0$ production measurements~\cite{miniboone_pi0,uboone_pi0,ArgoNeuT_pi0,K2K_pi0,SciBooNE_pi0,minerva_pi0,t2k_pi0}, which are sparse on argon targets and in the few-GeV regime.

To this end, we report the first double-differential cross section measurement of NC$\pi^0$ production in neutrino-argon scattering. The kinematics of the final state neutral pions are quantified by performing the measurement as a function of the $\pi^0$ momentum, $P_{\pi^0}$, and the cosine of the $\pi^0$ scattering angle with respect to the neutrino beam, $\cos\theta_{\pi^0}$. The signal definition includes events in which a neutrino of any flavor scatters via the neutral-current process and produces a single final state $\pi^0$ with $P_{\pi^0}<1.2$~GeV/c. The upper limit on the momentum restricts the measurement to regions of phase space with appreciable signal. Any hadronic final state not including an additional $\pi^0$ is included in the signal definition. 

In the same variables, single-differential measurements in terms of final states with and without protons are also reported. These use the signal definition outlined above but divide the semi-inclusive channel (``Xp") into final states containing no protons with kinetic energy above 35~MeV (``0p") and final states containing at least one proton with kinetic energy above 35~MeV (``Np"). Understanding the 0p and Np final states is particularly important for experiments employing liquid argon time projection chambers (LArTPCs)~\cite{LARTPC1,LARTPC2,LARTPC3,LARTPC4,LARTPC5}, which may utilize the presence of a gap between the neutrino and shower vertices to help differentiate electrons from photons~\cite{pelee,wc_elee,dl_elee,eLEE_PRL}. The tendency for no additional neutrino vertex activity in single-shower 0p events makes this topology particularly challenging and increases the NC$\pi^0$ background in these selections. This is especially important when testing BSM models, many of which predict single-shower final states without hadronic activity~\cite{BSM2,BSM3,BSM4,BSM6}. 

This work utilizes the data set collected with the MicroBooNE LArTPC detector~\cite{uboone_detector} and $6.4\times 10^{20}$ protons on target (POT) from the Booster Neutrino Beam (BNB)~\cite{MiniBooNEFlux}. The BNB is primarily composed of $\nu_\mu$ (93.7\%) with smaller $\bar{\nu}_\mu$ (5.8\%) and $\nu_e/\bar{\nu}_e$ (0.5\%) components. The MicroBooNE detector's TPC has 85 tonnes of liquid argon active mass and an array of 32 photomultiplier tubes (PMTs). Interactions that produce charged particles in the TPC create scintillation light and ionization electrons. The light is recorded by the PMTs which provides ns-scale timing for interactions~\cite{ns_timing}. The ionization electrons drift in a 273~V/cm electric field and induce charge on a set of three wire readout planes. Individual wire charge distributions are deconvolved from the detector response~\cite{tpc_signal_proc,sig_proc_1,sig_proc_2} and serve as inputs to the Wire-Cell topographical three-dimensional image processing algorithm~\cite{wc_reco}. This event reconstruction chain provides the basis for particle identification, calorimetry, and event selection~\cite{WC3D}. 

The Wire-Cell reconstruction identifies particle candidates by finding kinks in a cluster of activity in-time with the neutrino beam~\cite{WC3D}. Track and electromagnetic shower topologies are separated based on the amount of large-angle scattering, the proximity to additional isolated activity, and the width of the activity perpendicular to its trajectory. Candidate neutrino interaction vertices are formed concurrently alongside the hypothesized final state particles and their decay and scattering products based on the rate of deposited charge ($dQ/dx$), topology of the final state, and particle relationships~\cite{WC3D}. A final neutrino vertex is chosen by a $\texttt{SparseConvNet}$ neural network~\cite{SparseConvNet}. 

A shower's energy is reconstructed from its total deposited charge multiplied by a scale factor obtained from simulation that accounts for bias in charge reconstruction and the average recombination effect~\cite{wire-cell-uboone,charge_scale_1,charge_scale_2}.
An additional 0.95 scaling factor is applied to data based on previous $\pi^0$ mass calibration; this factor is not applied to the simulation~\cite{wc_elee}. The energy of tracks longer than 4~cm that stop within the active volume of the detector is calculated based on range using the NIST PSTAR database~\cite{pstar} with a correction for different particle masses. For all other tracks, the kinetic energy is calculated by converting their $dQ/dx$ to $dE/dx$ with an effective recombination model~\cite{WC3D,uboone_energy_cal}.

Neutral pions are reconstructed based on the identification of the two photons and their associated topological information. For events with more than two showers, the pair with the highest energy is used. The distance between the reconstructed neutrino vertex and $\pi^0$ vertex is used to separate primary pions from those produced in reinteractions outside the target nucleus. When NC events do not have additional hadronic activity to identify the $\pi^0$ vertex, the point on the each shower's primary axis that is closest to the opposite shower's primary axis is identified. The midpoint of the line connecting these two points is labeled as the displaced $\pi^0$ vertex and the direction of each shower is redefined with respect to that vertex~\cite{WC3D}. 

Monte Carlo (MC) simulations are used to train the boosted decision tree (BDT) event selection and estimate inputs to the data unfolding. The neutrino flux model utilizes MiniBooNE's $\texttt{Geant4}$-based simulation of the BNB~\cite{Geant4,MiniBooNEFlux}. Neutrino-argon interactions are simulated with the $\texttt{G18\_10a\_02\_11a}$ configuration of the $\texttt{GENIE v3.0.6}$ event generator~\cite{GENIE,GENIE2} that has been tuned to CC0$\pi$ data from T2K~\cite{T2KTuneData,uboonetune}. The tune has little impact on these measurements because it only affects charged-current (CC) quasi-elastic and meson-exchange-current events. Final state particles are propagated through a model of the detector using the $\texttt{Geant4}$ toolkit $\texttt{v4\_10\_3\_03c}$~\cite{Geant4} and $\texttt{LArSoft}$~\cite{larsoft} framework. The simulated TPC and PMT waveforms are overlaid on data taken without the neutrino beam to provide an accurate description of the cosmic ray backgrounds. These overlaid MC samples are processed with the Wire-Cell reconstruction in the same manner as real data.

The first step in selecting NC$\pi^0$ events is rejecting through-going and stopping cosmic ray muons with algorithms that match TPC-charge to PMT-light~\cite{cosmic_reject,wire-cell-uboone}. This forms the basis of the ``generic neutrino selection" which reduces cosmic backgrounds to about 15\% without imposing requirements on the nature of the neutrinos participating in the interactions. A BDT was then trained using the $\texttt{XGBoost}$ library~\cite{xgboost} on variables previously used to identify CC events~\cite{wc_elee} as well as additional reconstructed parameters designed to identify NC$\pi^0$ events. The training uses a signal enhanced sample of 40k events with the final BDT cut chosen based on maximising the product of the efficiency and purity. The selection achieves an efficiency of 35\% as estimated by the $\texttt{GENIE}$-based MicroBooNE MC. A total of 4971 candidate events fully-contained (FC) within the detector are selected when the BDT is applied to the data. This selection is estimated to have a 54\% purity for signal events. Figures illustrating the reconstruction quality, event selection efficiency, and measured distributions are presented in the Supplemental Material~\cite{supplemental}.

For the measurements of final states with and without protons, the selection is split into 0p and Np samples based on the presence of a reconstructed proton with kinetic energy greater than 35~MeV. This yields 1452 FC candidate Np NC$\pi^0$ data events and 3519 FC candidate 0p NC$\pi^0$ data events. The threshold is motivated by MicroBooNE's ability to detect tracks $>1$~cm in length, which corresponds to 35~MeV for protons, and is the energy at which the proton detection efficiency approaches 50\%~\cite{PRD}. Based on simulation, 92\% (54\%) of NC$\pi^0$ events that pass the Np (0p) selection satisfy the Np (0p) signal criteria.

The reconstructed $\pi^0$ momentum, $ P_{\pi^0}^{rec}$, and cosine of the reconstructed $\pi^0$ scattering angle, $\cos\theta_{\pi^0}^{rec}$, are calculated using the showers produced by the two photons associated with the $\pi^0$ decay. The opening angle of the showers, $\theta_{\gamma\gamma}$, and the energy of each shower, $E_{\gamma_1}$ and $E_{\gamma_2}$, is used to reconstruct $P_{\pi^0}^{rec}$  according to
\begin{equation}
    P_{\pi^0}^{rec} = m_{\pi^0}  \sqrt{\frac{2}{(1-\alpha^2)(1-\cos\theta_{\gamma\gamma})}-1} ,
\end{equation}
where $m_{\pi^0}=0.135$~GeV/c$^2$ is the $\pi^0$ mass~\cite{PDG}, and $\alpha = (E_{\gamma_1} - E_{\gamma_2})/(E_{\gamma_1} + E_{\gamma_2})$. The $P_{\pi^0}^{rec}$ resolution ranges from about 15\% at low momenta to about 40\% at high momenta. The $\pi^0$ scattering angle is calculated from the momentum of the two showers, $P_{\gamma_1}$ and $P_{\gamma_2}$, according to
\begin{equation}
    \cos\theta_{\pi^0}^{rec} = \frac{P_{\gamma}^z}{|\vec{P}_{\gamma}|},
\end{equation}
where $\vec{P}_{\gamma}=\vec{P}_{\gamma_1}+\vec{P}_{\gamma_2}$ and $P_{\gamma}^z$ is the component along the beam direction. The absolute $\cos\theta_{\pi^0}^{rec}$ resolution is around 0.1 but degrades at backwards angles for the 0p selection in large part due to less accurate neutrino vertex identification when additional vertex activity is not present.

%overview of the uncertainties
Systematic uncertainties on the reconstructed distributions are estimated with a total covariance matrix, $V^{\mathrm{sys}} =  V_{\mathrm{flux}} + V_{\mathrm{reint}} + V_{\mathrm{xs}} + V_{\mathrm{det}} + V_{\mathrm{MC}}^{\mathrm{stat}} + V_{\mathrm{dirt}} + V_{\mathrm{POT}} + V_{\mathrm{Target}}$, obtained by summing the covariance matrices calculated for each source of uncertainty. These are calculated simultaneously for 0p and Np events and thus include proper treatment of their correlations and the way each systematic migrates events between the two samples~\cite{PRD}.

The uncertainties on the BNB flux~\cite{MiniBooNEFlux} are contained in $V_{\mathrm{flux}}$, and the neutrino-argon interaction modeling uncertainties~\cite{uboonetune} are accounted for in $V_{\mathrm{xs}}$. These both contribute (5-15)\% uncertainty to the extracted cross sections and are similar in size to the data statistical uncertainty, except in some low count bins where the statistical uncertainty grows to (30-40)\%. Uncertainties on reinteractions outside the target nucleus are accounted for in $V_{\mathrm{reint}}$ using $\texttt{Geant4Reweight}$~\cite{geantrw}. These have little impact on the extracted results. The multi-sim technique~\cite{multisim} is used to calculate $V_{\mathrm{flux}}$, $V_{\mathrm{xs}}$, and $V_{\mathrm{reint}}$. Detector response uncertainties~\cite{detvar} are accounted for in $V_{\mathrm{det}}$ with a uni-sim approach. As in \cite{wc_elee,wc_1d_xs}, a single parameter is altered by $1\sigma$ and bootstrapping~\cite{boostrapping} is used to estimate the impact of this variation. Detector effects are the largest source of uncertainty on these measurements, usually contributing at the (10-25)\% level though rising to (30-60)\% at high $P_{\pi^0}$ and backwards $\cos\theta_{\pi^0}$. This is largely driven by significant detector uncertainties on the background prediction partially due to there being a lower number of background MC events available for bootstrapping. Also included are flat 50\%, 2\% and 1\% uncertainties on neutrino interactions outside the detector ($V_{\mathrm{dirt}}$), POT counting ($V_{\mathrm{POT}}$), and the number of target nuclei ($V_{\mathrm{Target}}$), respectively. Their impact on the total uncertainty is small. The Supplemental Material~\cite{supplemental} contains figures illustrating the contribution of each source of uncertainty to the total uncertainty on the extracted results.

Wiener-SVD unfolding~\cite{WSVD} is used to extract nominal flux-averaged cross section results~\cite{flux_uncertainty_rec}. The inputs for this method are the measurement $M$, the response matrix $R$ that describes the mapping between the true and reconstructed distributions predicted by the MC, and the reconstructed space covariance matrix $V=V^{\mathrm{sys}}+V^{\mathrm{stat}}$, where $V^{\mathrm{stat}}$ contains the data statistical uncertainty obtained following the combined Neyman-Pearson procedure~\cite{CNP}. The unfolding returns a regularized cross section and corresponding covariance matrix, $V_S$. An additional smearing matrix, $A_C$, capturing the bias induced by regularization is also obtained in the unfolding~\cite{WSVD}. Any prediction should be multiplied by $A_C$ when making a comparison to the data result. The extracted cross sections, $A_C$, and $V_S$ can be found in the Supplemental Material~\cite{supplemental}. 

The 0p and Np cross sections are extracted simultaneously following the formalism outlined in~\cite{PRD}, which accounts for the correlations between the 0p and Np channels during unfolding. This allows the number of true Np events in the 0p selection to be predicted based on the observation of the Np selection (and vice versa), thereby minimizing the overall dependence on the model. Alongside the FC sample, a smaller sample of 1467 events partially-contained (PC) within the detector are also collected and used in the unfolding. Due to larger uncertainties and lower event counts, these distributions have minimal impact on the results. Blockwise unfolding~\cite{PRD,GardinerXSecExtract} is also employed to obtain inter-variable correlations for the unfolded results.

Model inaccuracies can bias cross section measurements through inadequate estimations of the selection efficiency, background prediction, and the mapping between true and reconstructed variables. Data-driven model validation is employed to verify that the model, including its uncertainties, is sufficient for the unfolding. The model is deemed adequate if it can describe the data at the $2\sigma$ level. This is quantified with $\chi^2$ goodness-of-fit tests between measured and predicted distributions interpreted by using the number of degrees of freedom, $ndf$, which corresponds to the number of bins, to obtain $p$-values. To better expose relevant mismodeling, these tests utilize the conditional constraint formalism~\cite{cond_cov}. The conditional constraint leverages correlations between different channels and variables to update the model prediction and reduce the uncertainties on one distribution based on data observations in another distribution. The cross section extraction does not utilize these constraints, which are used strictly for model validation. This data-driven methodology is analogous to the model validation in other MicroBooNE analyses~\cite{wc_elee,PRD,PRL,wc_1d_xs,wc_3d_xs}. The model validation tests described below can be found in the Supplemental Material~\cite{supplemental}. 

Validating the modeling of $\pi^0$ kinematics starts by constraining the FC $P_{\pi^0}^{rec}$ prediction with the reconstructed neutrino energy distribution of the $\nu_\mu$CC selection from~\cite{wc_elee,PRD}. This constraint reduces correlated flux and detector uncertainties shared between NC$\pi^0$ and $\nu_\mu$CC events thereby better exposing the cross section modeling. This test is conducted on the distributions for the 0p, Np and Xp channels to evaluate each hadronic final state. Good agreement is observed, with $p$-values of 0.94, 0.84, and 0.80, for 0p, Np and Xp distributions, respectively. The same test is performed individually on all four angular slices used for the double-differential measurement and on the total reconstructed energy rather than $P_{\pi^0}$. The MC is able to describe the data within uncertainties in these tests with $p$-values close to one in all cases. 

To evaluate the modeling of the $\pi^0$ kinematics further, the FC $P_{\pi^0}^{rec}$ distribution is used to constrain the FC $\cos\theta_{\pi^0}^{rec}$ prediction. Correlations in the statistical uncertainties, arising from the fact that the constraining and constrained distributions utilize the same events, are estimated using a bootstrapping procedure~\cite{boostrapping,PRD}. These tests are applied to each hadronic final state and indicate that the data is described with uncertainties with $p$-values close to one in all cases. Alongside the tests on the $P_{\pi^0}^{rec}$ and reconstructed energy distributions in angular slices, this demonstrates that the overall model is sufficient for the extraction of the double-differential cross section as a function of  $\cos\theta_{\pi^0}$ and $P_{\pi^0}$.

The modeling of the proton kinematics is important for the simultaneous measurement of the 0p and Np NC$\pi^0$ cross sections. As such, the proton kinematics are evaluated with two separate constraints on the FC leading proton kinetic energy distribution. First, the reconstructed neutrino energy distribution from the $\nu_\mu$CC channel~\cite{wc_elee,PRD} is used; this results in a $p$-value of 0.90. Second, the FC $\pi^0$ kinematics are used; this constraint results in a $p$-value of 0.94. Together, with the validation of the $P_{\pi^0}$, $\cos \theta_{\pi^0}$, and reconstructed energy distributions for both the 0p and Np channels, these tests indicate that the model is sufficient for the simultaneous extraction of the 0p and Np cross sections. 

All aforementioned model validation tests are also applied to the PC distributions. These all yield $p$-values close to one. Additionally, 0p, Np, and Xp sidebands formed from a relaxed BDT selection criteria were studied to further verify proper background modeling. Good data to MC agreement is seen for these sidebands in the kinematic variables relevant to this analysis. The 0p normalization is well described by the model and the Np normalization slightly overestimated, but still within $1\sigma$. These studies can be found in the Supplemental Material~\cite{supplemental}.

\begin{figure}
\centering
  \begin{subfigure}[t]{0.9\linewidth}
\includegraphics[width=\linewidth]{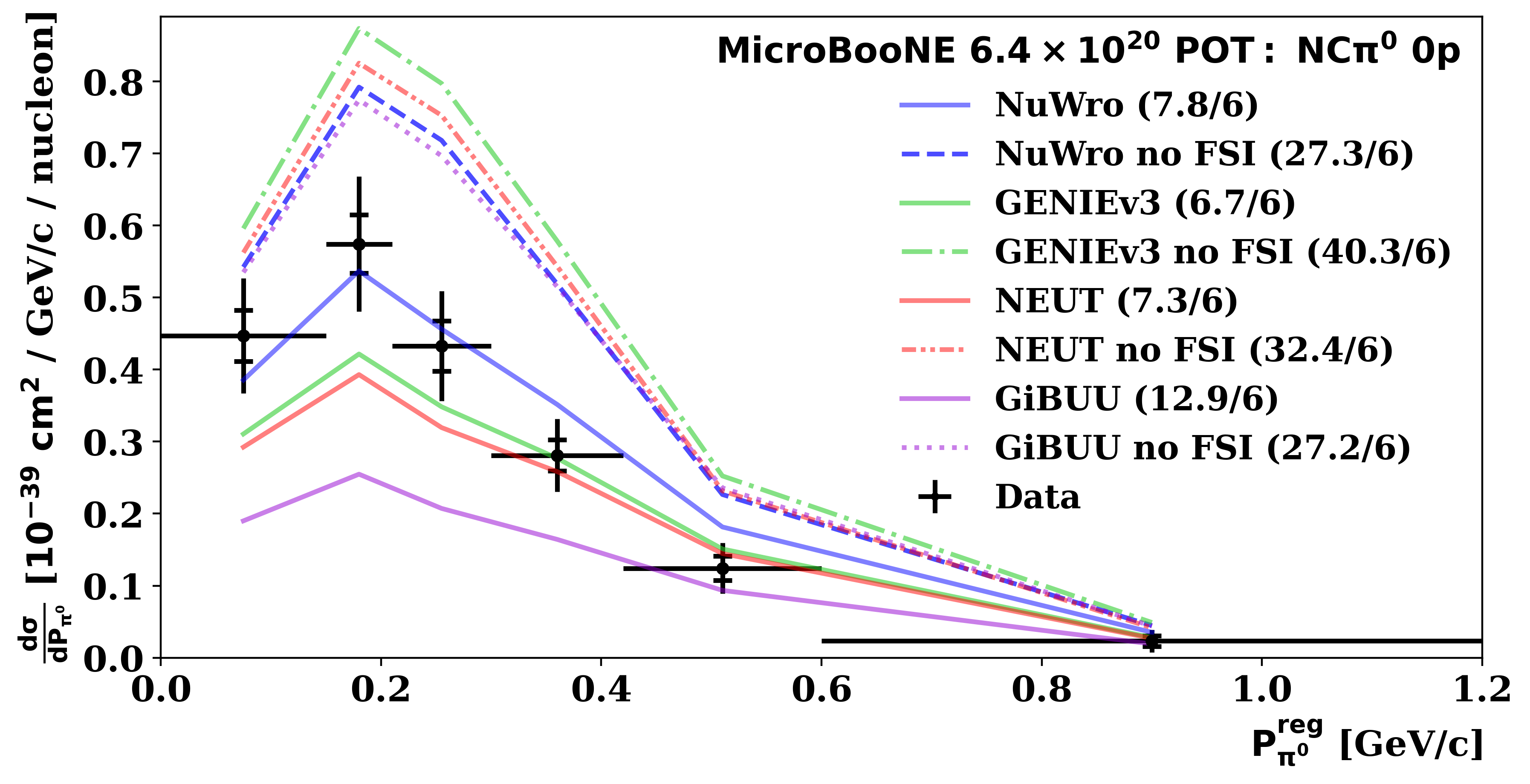}
  \put(-196pt,100pt){(a)} 
  \end{subfigure}
  \begin{subfigure}[t]{0.9\linewidth}
\includegraphics[width=\linewidth]{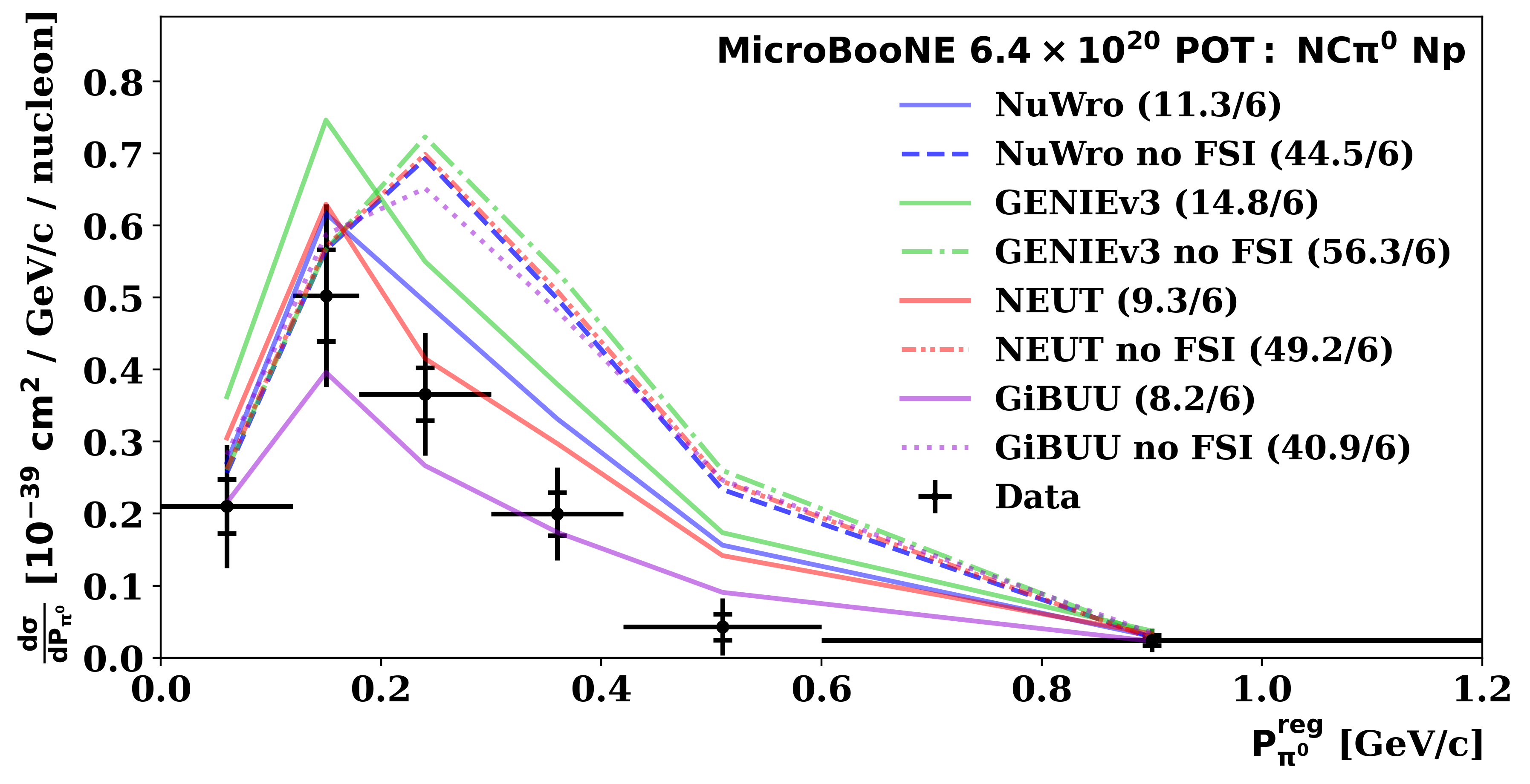}
  \put(-196pt,100pt){(b)} 
  \end{subfigure} 
  \caption{Unfolded 0p (a) and Np (b) $P_{\pi^0}$ differential cross sections. The black inner (outer) error bars on the data points represent the statistical (total) uncertainties on the extracted cross section corresponding to the square root of the diagonal elements of the extracted covariance matrix. Generator predictions are indicated by the colored lines with corresponding $\chi^2/ndf$ values displayed in the legend. The $\bf{^{reg}}$ superscript  indicates the results are regularized and predictions are smeared with $A_C$ to account for any bias.}
\label{mom_0pNp}
\end{figure}

\begin{figure}
\centering
  \begin{subfigure}[t]{0.9\linewidth}
\includegraphics[width=\linewidth]{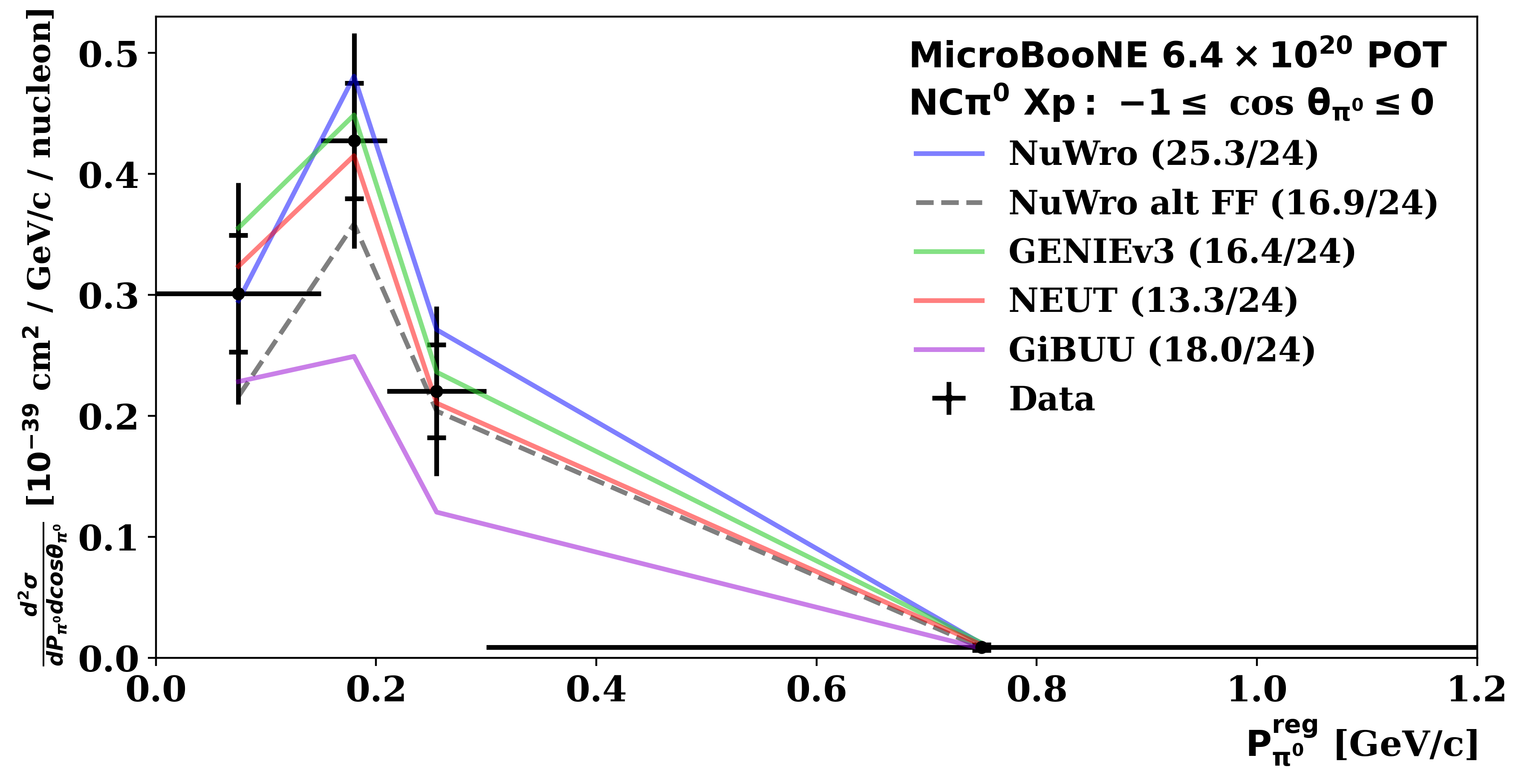}
  \put(-196pt,100pt){(a)} 
  \end{subfigure}
 \begin{subfigure}[t]{0.9\linewidth}
  \includegraphics[width=\linewidth]{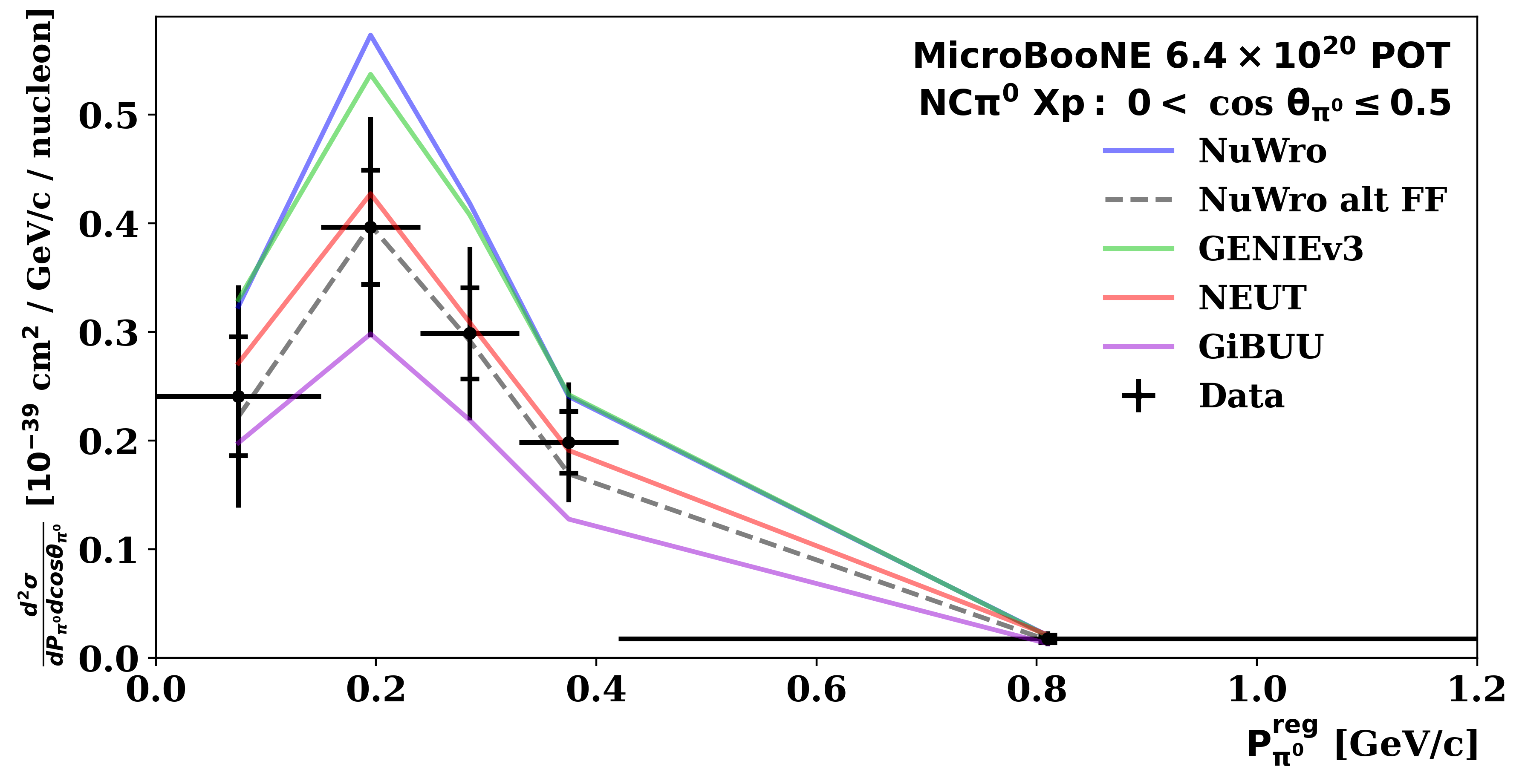}
  \put(-196pt,100pt){(b)}
  \end{subfigure}
    \begin{subfigure}[t]{0.9\linewidth}
\includegraphics[width=\linewidth]{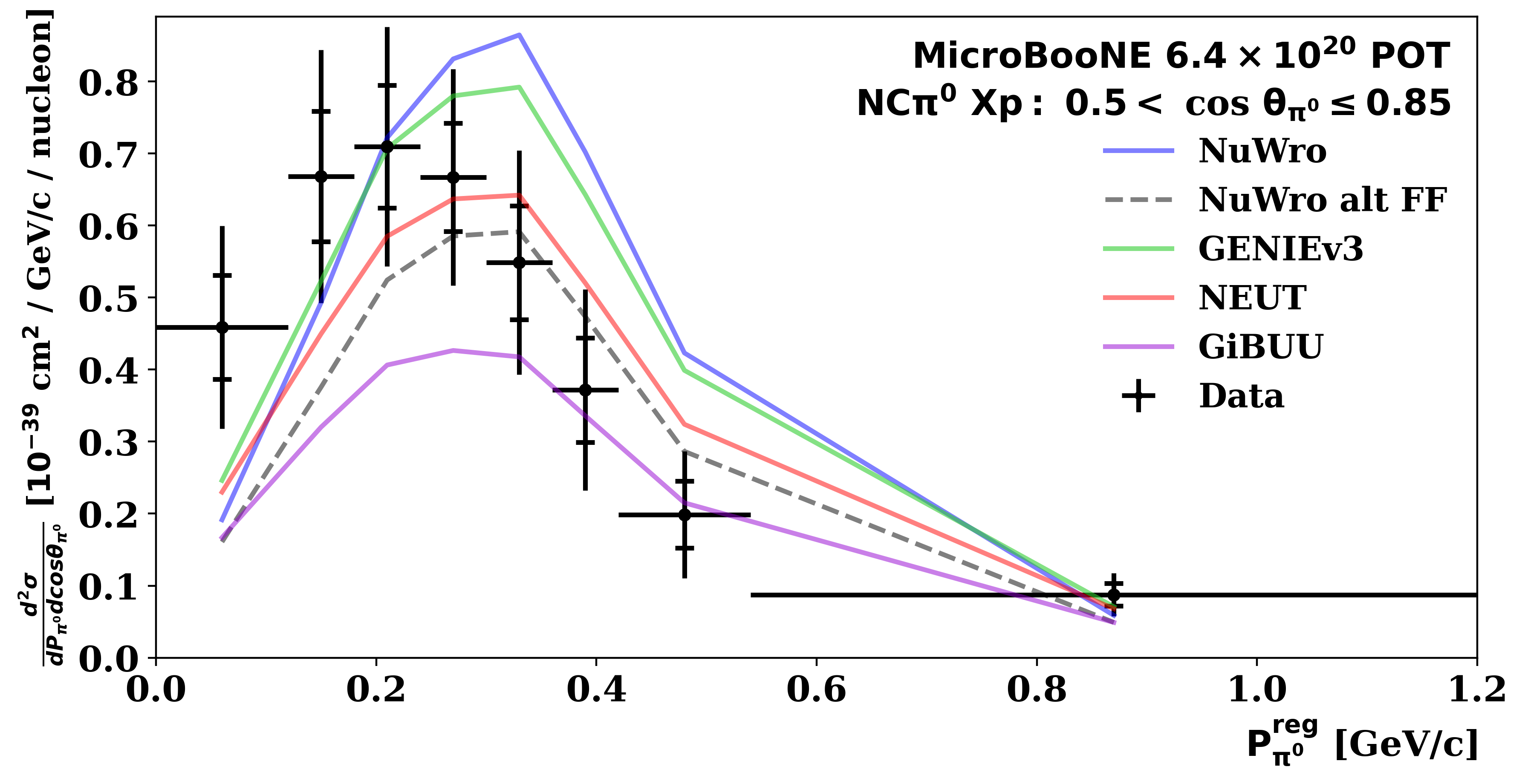}
  \put(-196pt,100pt){(c)}
  \end{subfigure}
 \begin{subfigure}[t]{0.9\linewidth}
  \includegraphics[width=\linewidth]{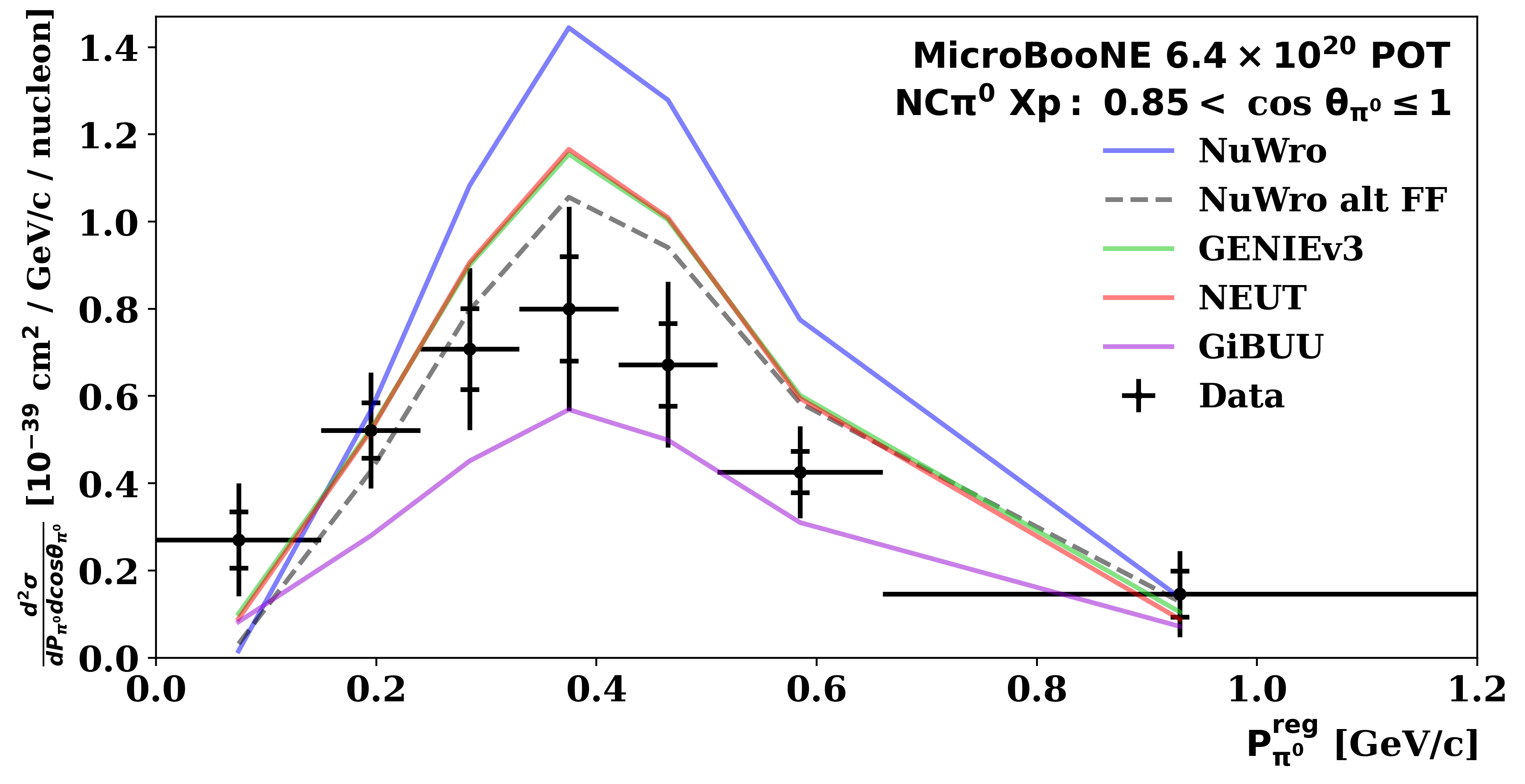}
  \put(-196pt,100pt){(d)}
  \end{subfigure}
  \caption{Same as Fig.~\ref{mom_0pNp}, but for the $\cos\theta_{\pi^0}$ and $P_{\pi^0}$ double-differential Xp cross section result. Each subfigure shows a different $\cos\theta_{\pi^0}$ angular region, with the $\chi^2/ndf$ calculated across all bins displayed in the legend of (a).}
\label{cosmom}
\end{figure}

The extracted cross section results are compared to event generator predictions from $\texttt{GENIE v3.0.6 G18\_10a\_02\_11a}$ ($\texttt{GENIEv3}$)~\cite{GENIE}, $\texttt{NuWro 21.02}$ ($\texttt{NuWro}$)~\cite{nuwro}, $\texttt{GiBUU 2023}$ ($\texttt{GiBUU}$)~\cite{gibuu2}, and $\texttt{NEUT 5.4.0.1}$ ($\texttt{NEUT}$)~\cite{neut}. To demonstrate the utility of these measurements, these comparisons include predictions which modify the FSI experienced by the outgoing particles, or the form factors describing the neutrino-nucleon interaction. Generator predictions were processed with the $\texttt{NUISANCE}$ framework~\cite{NUISANCE}, do not include theoretical uncertainties, and are smeared with the $A_C$ obtained from unfolding. Agreement with the data is quantified by $\chi^2$ values calculated with uncertainties according to $V_S$. 

The simultaneously extracted 0p and Np $P_{\pi^0}$ differential cross sections are shown in Fig.~\ref{mom_0pNp} alongside generator predictions with and without FSI. Compared to the ``no FSI" predictions, the predictions with FSI reduce the cross section, shift the peak of the $P_{\pi^0}$ distribution towards lower values resulting in a sharper drop just beyond the peak, and are favored by the data. This is unsurprising as similar features are well established in measurements of photoproduction of pions on nuclear targets~\cite{photo_pi} where, despite involving different probes, the FSI are identical to neutrino scattering. The shift towards smaller $P_{\pi^0}$ is less prominent for 0p, possibly due to the fact that reinteractions of the $\pi^0$ may also knock out protons. This redistributes events from 0p to Np and further reduces the 0p cross section. Nevertheless, the predictions with FSI still overestimate the measured Np NC$\pi^0$ cross section, particularly around the 0.2-0.5~GeV/c momentum range. The exception to this is $\texttt{GiBUU}$, which slightly underestimates the Np channel and strongly underestimates the 0p channel. This observation is interesting given that $\texttt{GiBUU}$ shows a better description of other MicroBooNE 0p measurements on the $\nu_\mu$CC channel~\cite{PRL,PRD} than other generators do. Its low normalization here points towards important subtleties in the treatment of FSI between nucleons, resonances, and mesons~\cite{dytman_transparency}. The 0.2-0.5~GeV/c momentum range is strongly impacted by FSI, suggesting that refinements to FSI modeling may enable a better description of this data.

Figure~\ref{cosmom} shows the unfolded double-differential Xp cross section as a function of $P_{\pi^0}$ for specific $\cos\theta_{\pi^0}$ regions. Generator predictions are also shown. $\texttt{NEUT}$ describes the data best followed by $\texttt{GENIEv3}$ and $\texttt{GiBUU}$. $\texttt{NuWro}$ has the worst description of the data due to a consistent overestimation of the cross section, but its performance is significantly improved if, rather than the default dipole parameterization~\cite{dipole}, an alternative set of axial form factors with steeper $Q^2$ dependence~\cite{olga} is utilized. The latter prediction, labeled $\texttt{NuWro~alt~FF}$, shows better normalization agreement. This observation is interesting because when the analogous form factors are compared to ANL CC$\pi^+$ deuterium bubble chamber data~\cite{anl1,anl2}, which do not contain significant nuclear effects, the dipole prediction also overestimates the data and better agreement is seen for the steeper $Q^2$ dependence.

For $\cos\theta_{\pi^0}<0$, generator description of the data is overall sufficient, though $\texttt{GiBUU}$ does show some underprediction of the cross section. For $\cos\theta_{\pi^0}>0$, $\texttt{NEUT}$ and $\texttt{NuWro~alt~FF}$ perform best but their description of the data is not as good as it was for more backwards angles. $\texttt{GENIEv3}$ and $\texttt{NuWro}$ begin to overpredict the data in the $0<\cos\theta_{\pi^0}<0.5$ region, especially at low-to-moderate momenta, and $\texttt{GiBUU}$ still underestimates it. In the $0.5<\cos\theta_{\pi^0}<0.85$ region, all generators predict that the peak in the momentum distribution occurs at higher values than seen in data. In the forward angle $\cos\theta_{\pi^0}>0.85$ region, all generators underestimate the cross section at low momentum and overestimate it around and just beyond the peak of the distribution. The exception is $\texttt{GiBUU}$, which consistently underestimates the cross section instead. These discrepancies at forward angles, and the qualitative difference between $\texttt{GiBUU}$ and other generators in the forward direction, could possibly be due to FSI, which shifts the momentum distribution towards lower values, or the modeling of coherent pion production, which is included in $\texttt{NEUT}$, $\texttt{NuWro}$, and $\texttt{GENIE}$, but not $\texttt{GiBUU}$. It may also suggest a need for different $Q^2$ dependence.

In addition to the measurements described above, simultaneously extracted 0p and Np $\cos\theta_{\pi^0}$ differential cross sections are presented in the Supplemental Material~\cite{supplemental}. Semi-inclusive Xp single-differential cross sections in $P_{\pi^0}$ and $\cos\theta_{\pi^0}$ are also included.

In summary, we report the first double-differential cross section measurements of neutral-current $\pi^0$ production in neutrino-argon scattering. Single-differential measurements in terms of final states with and without protons are also reported. These measurements are performed with a boosted decision tree based event selection and, after extensive model validation to ensure unbiased unfolding, are extracted with the Wiener-SVD method. Commonly used neutrino event generators overestimate the measured NC$\pi^0$ cross section, especially for $\pi^0$ momentum around 0.2-0.5~GeV/c, at forward scattering angles, or when a proton is present in the final state. The exception to this is $\texttt{GiBUU}$, which instead underestimates the cross section. The NC$\pi^0$ channel is a critical background in oscillation analyses and BSM searches, and these results are a step towards improving the modeling of this under-characterized channel.
\begin{acknowledgments}
This document was prepared by the MicroBooNE collaboration using the resources of the Fermi National Accelerator Laboratory (Fermilab), a U.S. Department of Energy, Office of Science, HEP User Facility. Fermilab is managed by Fermi Research Alliance, LLC (FRA), acting under Contract No. DE-AC02-07CH11359.  MicroBooNE is supported by the following: the U.S. Department of Energy, Office of Science, Offices of High Energy Physics and Nuclear Physics; the U.S. National Science Foundation; the Swiss National Science Foundation; the Science and Technology Facilities Council (STFC), part of the United Kingdom Research and Innovation; the Royal Society (United Kingdom); and the UK Research and Innovation (UKRI) Future Leaders Fellowship. Additional support for the laser calibration system and cosmic ray tagger was provided by the Albert Einstein Center for Fundamental Physics, Bern, Switzerland. We also acknowledge the contributions of technical and scientific staff to the design, construction, and operation of the MicroBooNE detector as well as the contributions of past collaborators to the development of MicroBooNE analyses, without whom this work would not have been possible. For the purpose of open access, the authors have applied a Creative Commons Attribution (CC BY) public copyright license to any Author Accepted Manuscript version arising from this submission.
\end{acknowledgments}

\bibliography{prl.bib}% Produces the bibliography via BibTeX.

\end{document}